\documentclass[final,3p,twocolumn]{elsarticle}

\usepackage{graphicx}
\usepackage{nicefrac}
\usepackage{amsmath}
\usepackage{cuted}
\usepackage{slashed}

\usepackage[usenames,dvipsnames]{xcolor}

\newcommand{\Slash}[1]{\ooalign{\hfil/\hfil\crcr$#1$}}

\newcommand{\bracket}[1]{\left( #1 \right)}

\begin{document}

\title{Electromagnetic and strong isospin breaking in light meson masses}

\makeatletter
\let\cat@comma@active\@empty
\makeatother

\author[morelia]{\'Angel S. Miramontes}
\ead{angel-aml@hotmail.com}
\address[morelia]{Instituto de F\'isica y Matem\'aticas, Universidad Michoacana de San Nicol\'as de Hidalgo, Morelia, Michoac\'an 58040, Mexico}
\author[graz]{Reinhard Alkofer}
\ead{reinhard.alkofer@uni-graz.at}
\address[graz]{Institute of Physics, University of Graz, NAWI Graz, Universit\"atsplatz 5, 8010 Graz, Austria}
\author[giessen]{Christian S. Fischer}
\ead{christian.fischer@uni-giessen.de}
\address[giessen]{Institut f\"ur Theoretische Physik, Justus-Liebig-Universit\"at Giessen, 35392 Giessen, Germany;\\
	Helmholtz Forschungsakademie Hessen f\"ur FAIR (HFHF)
          GSI Helmholtzzentrum f\"ur Schwerionenforschung, Campus Gie{\ss}en, 35392 Gie{\ss}en, Germany}
\author[sal]{H\`elios Sanchis-Alepuz}
\ead{helios.sanchis-alepuz@silicon-austria.com}
\address[sal]{Silicon Austria Labs GmbH, Inffeldgasse 33, 8010 Graz, Austria}

\date{\today}

\begin{abstract}
We study electromagnetic as well as strong isospin breaking effects in the isospin mass splittings
of light pseudoscalar and vector mesons. To this end we employ a coupled system of 
quark Dyson-Schwinger and meson Bethe-Salpeter equations whose interaction 
kernels contain gluon, pion and photon exchange interactions. 
In bound states, QCD-induced isospin breaking is manifest on different 
levels. On the one hand, a different explicit up- and down-quark mass directly affects the 
propagators of the constituent quarks. On the other hand, it leads to different interaction kernels 
within the isospin multiplets. In addition, electromagnetic isospin breaking is induced via a  
photon exchange diagram. Using the kaon iso-doublet and the charged pion masses
as input to determine the up, down and strange quark masses we find
for the pion, kaon and rho meson mass splittings different patterns each. In particular, our 
results provide evidence that the effects from two sources of mass splittings, the different quark masses and
the different quark charges, do not add up linearly.
\end{abstract}

\maketitle

\section{\label{sec:intro}Introduction}

The approach to QCD via functional methods, most prominently with 
Dyson-Schwinger equations, Bethe-Salpeter equations and/or the 
functional renormalisation group has reached a degree of sophistication that allows to systematically study 
sub-leading effects in the hadron spectrum. Arguably the most important of these effects is the breaking of
isospin symmetry. It plays an important role in baryon physics, notably in the mass splitting between the
proton and the neutron which is related to fundamental questions like the existence and stability of 
ordinary matter in our universe. It also plays an important role in meson physics allowing for a range
of interesting and important isospin violating decays such as the one of the exotic
X(3872). Furthermore, its inclusion is mandatory for precision determinations of hadronic contributions 
to the anomalous magnetic moment of the muon \cite{Aoyama:2020ynm,Borsanyi:2020mff},
and of the CKM matrix elements $V_{ud}$ and $V_{us}$ (see, {\it e.g.},
\cite{Seng:2021nar} for a recent update on the respective analysis of $\pi$ and $K$ decays in view of the
apparent 3$\sigma$ violation of the first-row CKM unitarity condition). Thus, a better understanding
of isospin breaking effects, besides being interesting in its own right, might play a significant r\^ole 
in the detection of new physics via high-precision experiments.

The mass difference of the charged to the neutral pions have been analysed already before the 
advent of QCD \cite{Das:1967it}  with the conclusion that the observed pion mass difference does not arise
from the mass difference of the two light quark flavours but from the electromagnetic interaction.
Based on this, when considering three flavours, one arrives at the conjecture for the electromagnetically
induced kaon mass difference
\begin{equation}
\left( m^2_{K^\pm}- m^2_{K^0} \right)_{\mathrm em} = m^2_{\pi^\pm}- m^2_{\pi^0}
\end{equation}
which is also known as Dashen's theorem \cite{Dashen:1969eg}.
However, this relation is already significantly violated beyond leading order Chiral Perturbation Theory ($\chi$PT) 
\cite{Bijnens:1996kk,Amoros:2001cp}, the next-to-leading order providing 
approximately a factor of two   
\begin{equation}
\label{eq:DashenCorr}
\left( m^2_{K^\pm}- m^2_{K^0} \right)_{\mathrm em} = (1.84\pm0.25) 
\left( m^2_{\pi^\pm}- m^2_{\pi^0} \right) \, .
\end{equation}
From this one may conclude that the electromagnetic interaction alone would imply that the charged 
kaon is approximately 2 MeV heavier than the neutral one. Note, however, that the observed
mass ordering is opposite \cite{ParticleDataGroup:2020ssz}
\begin{equation}
m_{K^0}  - m_{K^\pm} = (3.934 \pm 0.0020) \, {\mathrm{MeV}}~,
\end{equation}
the sign and the magnitude of the kaon mass difference being mostly due to the light quark
mass difference.

In the past years, lattice QCD calculations with QED included have 
delivered detailed analyses of isospin breaking effects and determined successfully and with increasing precision the resulting masses of the up and down quarks, see
the FLAG-report \cite{Aoki:2021kgd} for an overview and \cite{Horsley:2015vla,MILC:2018ddw,Fodor:2016bgu,Giusti:2017dmp,Yong:2021qum}
as well as references therein.

Isospin breaking effects have been studied in functional methods already in the pioneering work of 
Jain and Munczek \cite{Jain:1993qh} using a simple but efficient truncation of the quark-gluon 
interaction in terms of  one-gluon exchange. 
An improved ladder approximation in Dyson-Schwinger and Bethe-Salpeter equations have been
used in refs.\ \cite{Harada:2004qn,Harada:2005ru} to calculate the pion mass difference.
Since then, the framework has been developed to 
include the full momentum dependent physics from all primitively divergent Green's functions leading to spectra in qualitative and in most channels even quantitative agreement with experiment, see 
\cite{Eichmann:2016yit} for a review. From a physics perspective,
a particularly interesting contribution to the quark-gluon interaction stems from four-quark functions, 
which can be parameterised in terms of (off-shell) meson contributions. These reappear, as 
expected from general properties of QCD and, in particular, from quark-hadron duality, in the 
self-energies and interaction kernels of Bethe-Salpeter equations as meson loops and exchanges, 
which can be interpreted as meson cloud effects. As we  will see in the course of this work, these meson cloud effects play an important part in the systematics of strong breaking of isospin due to the mass 
difference of the up- and down-quark. The second, in many cases competing, effect is isospin breaking 
due to the charge difference of the up- and down-quark. Diagrammatically, this QED effect
appears in the form of photon loop and exchange diagrams which have to be added to the self-energies 
and interaction kernels.

In this work, we study contributions of strong and electromagnetic isospin breaking effects on the
mass splitting of the isospin multiplets of light pseudoscalar and vector mesons. Based on previous
work on the pion electromagnetic  form factor \cite{Miramontes:2021xgn}, our truncation scheme 
takes explicitly into account that the non-perturbative  interaction between light quarks becomes 
flavour dependent away from the isospin limit. We summarise technical
details of our scheme in section \ref{sec2} and discuss our results for the mass splittings and 
the associated up- and down-quark masses in section \ref{sec:results}. In section \ref{sec:outlook} 
we conclude with an outlook to future applications. 

\begin{figure*}[th]
	\centerline{
		\includegraphics[width=\textwidth]{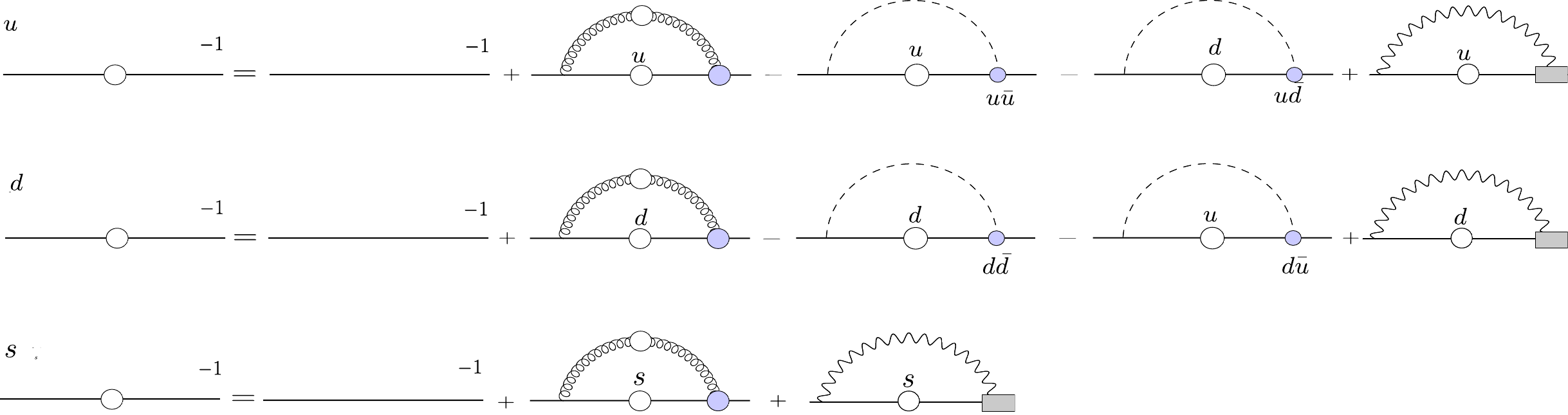}}
	\caption{Coupled system of DSEs for up and down quarks (upper two equations) including explicit pion backreaction diagrams 
		and a loop including a photon. Gluons are represented by curly lines, mesons by dashed lines and 
		photons by wiggly lines. In each diagram the left vertex is bare (but renormalised) and the vertex 
		on the right is dressed. Kaon and eta backreaction diagrams are neglected (see main text). Correspondingly,
		the DSE for the strange quark (lower equation) contains gluon and photon diagrams only. 
		}
	\label{fig:DSE}
\end{figure*}

\section{Formalism}\label{sec2}
In this section we discuss our heuristic approach to isospin breaking using the formalism of 
Dyson-Schwinger (DSE)  and Bethe-Salpeter (BSE) equations. We briefly summarise the key elements 
of the DSE/BSE approach (for more details 
see e.g. \cite{Cloet:2013jya, Eichmann:2016yit,Huber:2018ned,Sanchis-Alepuz:2017jjd}) and introduce 
the two mechanisms  for isospin breaking that we explore in this paper.

Formally, the quark-DSE for a quark with flavour $f$ is given by 
\begin{align}\label{eq:DSE}
	S^{-1}_f(p) =& Z_2 \,S^{-1}_{0,f}(p) - Z_{1f} \,g^2 \,C_F \times \\
	&\times\int \!\! \frac{d^4q}{(2\pi)^4}\, i\gamma^\mu S_f(q) \,\Gamma_{\mathrm{qg},f}^\nu(q,p) \,D^{\mu\nu}(k)\;,\nonumber
\end{align}
where $Z_2$ and $Z_{1f}=Z_g \,Z_2 \,Z_3^{1/2}$ are (flavour-dependent) renormalisation constants
for the quark  propagator and quark-gluon vertex, and $C_F=4/3$ is the colour Casimir for $N_c=3$.
The inverse tree-level propagator  is given by $S^{-1}_{0,f}(p) = i\Slash{p}+Z_m \,m_f$, where
$m_f$ is the renormalized quark mass from the QCD action. The dressed quark-gluon vertex $\Gamma_{\mathrm{qg},f}$
has a very rich structure including gluonic but also effective meson exchange contributions detailed below.

The dressed quark propagators, {\it i.e.}, the solutions of the DSE \eqref{eq:DSE}, will be of the form
\begin{align}\label{QuarkProp}
S_f(p) = \frac 1 {A_f(p^2)} \frac {-i \Slash{p} +M_f(p^2)}{p^2+M^2_f(p^2)}
\end{align}
which, for each flavour, is determined by two dressing functions, $A_f(p^2)$ and $M_f(p^2)$. The latter are the 
dynamically generated mass functions, and we will use below $M_f(p^2=0)$ to demonstrate that in the solutions of the 
DSE \eqref{eq:DSE} the isospin breaking effects due to the current mass differences are enhanced.

In the BSE framework, bound states of a quark and an antiquark can be described by 
Bethe-Salpeter amplitudes $\Gamma$ 
which are obtained as solutions of a homogeneous BSE, 
\begin{flalign}
\label{eq:homogeneousBSE}
\Gamma_{fg}\bracket{p,P}&= \sum_{f^\prime g^\prime} \int_q K_{fg}^{f^\prime g^\prime}\bracket{P,p,q}
\Bigl( S_{f^\prime} \bracket{k_1}  \times \nonumber\\
&\qquad \times
\Gamma_{f^\prime g^\prime} \bracket{q,P}S_{g^\prime} \bracket{k_2}\Bigr)~,
\end{flalign}
with $P$ the total meson momentum, $p$ is the relative momentum between quark and antiquark, 
$q$ is an internal relative  momentum which is integrated over. 
For pseudoscalar and vector mesons, the Dirac part of the Bethe-Salpeter amplitude 
$\Gamma_{fg}$ for a meson with constituents of flavour $f,g$
can be expanded in a tensorial basis with four and eight  elements, respectively. 
The interaction kernel $K$ encodes all  possible interactions between quarks and anti-quarks, 
and the quark propagators $S_{f}(p)$ are given by the solution of the quark DSE in Eq.~\eqref{eq:DSE}.

\begin{figure*}[t!]
	\centerline{
		\includegraphics[width=\textwidth]{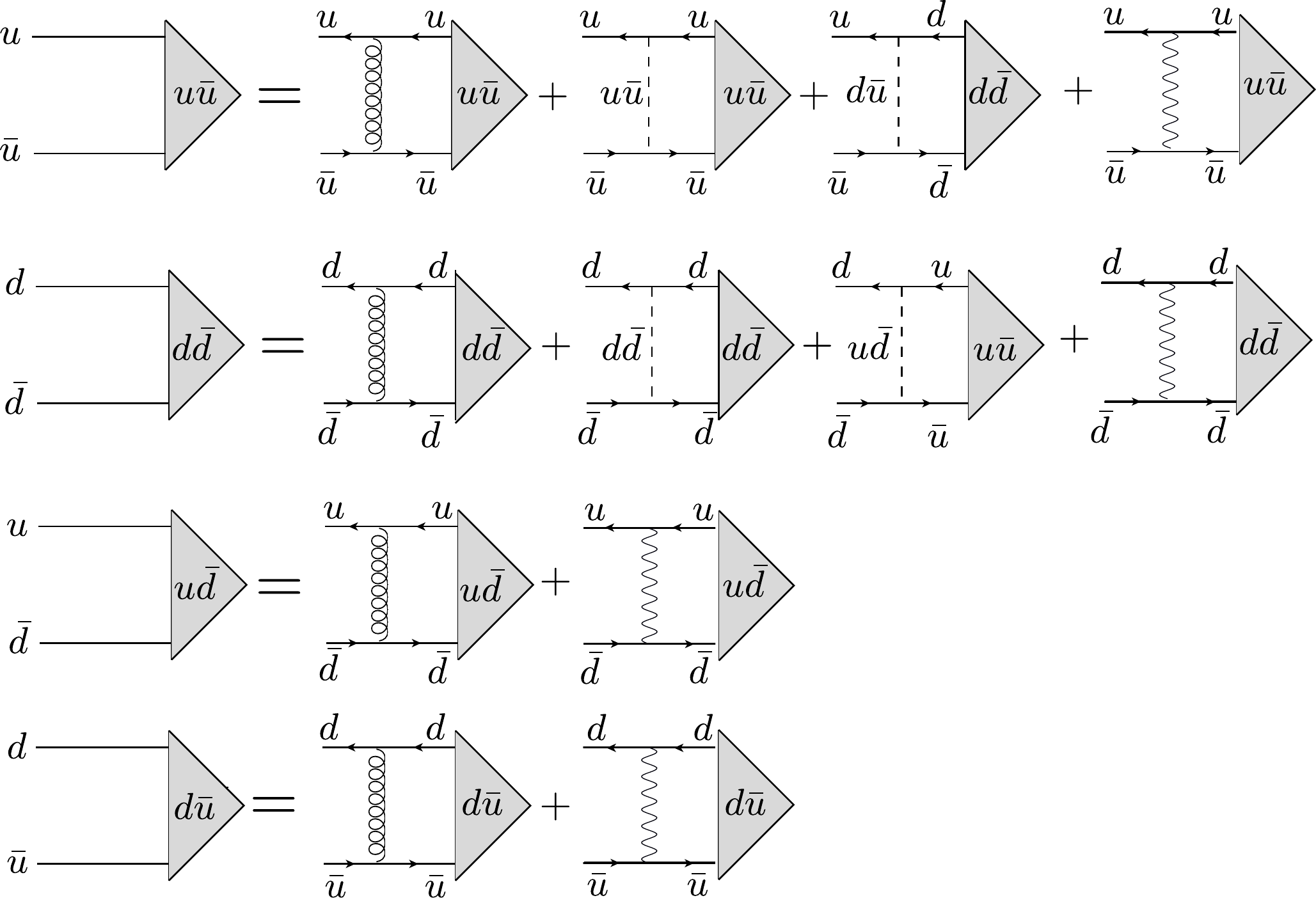}}
	\caption{Coupled system of BSEs for the isospin triplet including explicit meson and photon exchange diagrams.
		     The corresponding BSEs for the kaons are similar in structure as the ones for the charged pions. }
	\label{fig:BSE}
\end{figure*}

In this work we consider three types of contributions to the quark self-energy and the BSE interaction 
kernel: 
(i) a  flavour-blind dressed quark-antiquark gluon exchange that provides the necessary interaction 
strength to form mesonic bound states, 
(ii) a meson-exchange mechanism that provides flavour mixing and, 
(iii) a dressed photon exchange describing electromagnetic interactions between the quark and 
the antiquark. 
The latter is responsible for the electromagnetically induced isospin breaking. These contributions 
are shown explicitly in Figs.~\ref{fig:DSE} and \ref{fig:BSE}. In the following, we explain each contribution
in turn.

\subsection{\label{sec:explicit}Explicit Isospin Breaking and Flavour Mixing}

Let us consider first the non-electromagnetic contributions to the quark self-energy and the BSE interaction kernel. 
As indicated above and shown in Figs.~\ref{fig:DSE} and \ref{fig:BSE} the strongly-interacting part consists of a
gluon loop/exchange term and meson loop/exchange terms.

For the gluon term we use the rainbow-ladder truncation of the quark-gluon vertex, whereby the effect of (the leading term of)
the dressed quark-gluon vertex $\Gamma_\mathrm{qg,f}^\nu(q,p) = i \gamma^\nu \Gamma(k^2)$ and the Landau-gauge dressed
gluon propagator $D^{\mu\nu}(k) = T^{\mu\nu} Z(k^2)/k^2$ with transverse projector $T$ is combined into a effective 
coupling that only depends on the gluon momentum $k$. 
\begin{equation}\label{alpha-eff}
	\alpha(k^2) = \frac{Z_{1f}}{Z_2^2}\,\frac{g^2}{4\pi}\,Z(k^2)\,\Gamma(k^2)\,.
\end{equation}
For the effective coupling we use the Maris-Tandy model \cite{Maris:1997tm, Maris:1999nt}
\begin{flalign}\label{eq:MTmodel}
\alpha_{\text{QCD}}(k^2)=&
 \pi\eta^7\left(\frac{k^2}{\Lambda^2}\right)^2
e^{-\eta^2\frac{k^2}{\Lambda^2}}\nonumber\\
&+\frac{2\pi\gamma_m
\big(1-e^{-k^2/\Lambda_{t}^2}\big)}{\textnormal{ln}[e^2-1+(1+k^2/\Lambda_{QCD}
^2)^2]}~,
\end{flalign}
where the Gaussian term provides binding strength and enables dynamical chiral symmetry breaking; the second term 
reproduces the one-loop QCD behavior of the quark propagator in the ultraviolet when used in the quark DSE. 
In this model the scale $\Lambda_t=1$~GeV is introduced for technical reasons. For the anomalous dimension we use $\gamma_m=12/(11N_C-2N_f)=12/25$ with $N_f=4$ flavours and $N_c=3$ colours. For the QCD scale we use 
$\Lambda_{QCD}=0.234$ GeV. The parameters $\eta$ and $\Lambda$ are discussed below.

Note that such a truncation of the gluon-interaction guarantees that the axial-vector WTI (axWTI) is respected and, 
hence, that the pion is massless in the chiral limit.

A gluon exchange as interaction mechanism is flavour diagonal. In order to introduce flavour mixing we complement 
the interaction kernel with a meson exchange mechanism (see Fig.~\ref{fig:BSE}).
The BSE kernel, defined in \cite{Fischer:2007ze, Fischer:2008sp} and representing the exchange of a pion with 
Bethe-Salpeter amplitude $\Gamma_{\pi}$ and propagator $D_\pi(p) = 1/(p^2+m_\pi^2)$, reads
\begin{flalign}
&K^{(t)~ut}_{rs}(q,p;P) =\nonumber\\
 &~\frac{C}{4} [\Gamma_{\pi}^j]_{ru} \left(\frac{p + q - P}{2}; p - q \right) [Z_2 \gamma^5]_{ts} D_{\pi}(p - q) \nonumber \\ \nonumber
 &+\frac{C}{4} [\Gamma_{\pi}^j]_{ru} \left(\frac{p + q - P}{2}; q - p \right) [Z_2 \gamma^5]_{ts} D_{\pi}(p - q) \\ \nonumber
 &+\frac{C}{4} [Z_2 \gamma^5]_{ru} [\Gamma_{\pi}^j]_{ts} \left(\frac{p + q + P}{2}; p - q \right) D_{\pi}(p - q) \\
 &+\frac{C}{4} [Z_2 \gamma^5]_{ru} [\Gamma_{\pi}^j]_{ts} \left(\frac{p + q + P}{2}; q - p \right) D_{\pi}(p - q).\label{eq:BSEkernel_tchannel}
\end{flalign}
The quark DSE, correspondingly, 
is extended with the addition of further loop diagrams (third and fourth terms in the quark DSEs in Fig.~\ref{fig:DSE}).
As discussed in detail in \cite{Fischer:2007ze,Fischer:2008sp}, the presence of these loops can be motivated and derived 
from the underlying Dyson-Schwinger equation of the quark-gluon vertex, where meson exchange diagrams appear as approximation
of a momentum dependent four-Fermi interaction. In a complete treatment of these effects, we would need to include 
meson diagrams involving pions, kaons and in the DSE for the strange quark even the $\eta/\eta'$ mesons. However, 
already pion backraction effects are subleading on the 10-20 percent level compared to gluon effects
\cite{Fischer:2007ze,Fischer:2008sp}. Since kaon and $\eta/\eta'$ effects are further suppressed by factors of 
$m_\pi^2/m_{K,\eta,\eta'}^2$ it seems safe to neglect these in a first approach. A similar argument applies to potential
contributions from mesons with other quantum numbers.

In contrast to the gluon exchange, the meson exchange mechanism does not immediately preserve the axWTI and can, potentially, 
violate chiral symmetry. It was shown in \cite{Fischer:2007ze,Fischer:2008sp} how an axWTI-preserving meson-exchange kernel 
can be constructed. We use this kernel, in particular, we set $C=-3/2$,
 in this work. Note that consequently the meson exchange kernels do not introduce any additional 
parameters.

The resulting diagrams in the BSEs are shown in Fig.~\ref{fig:BSE}. It is interesting to note that due to the flavour 
structure, the meson exchange diagrams only appear in the flavour diagonal BSEs for the $u\bar{u}$ and $d\bar{d}$ components.
Corresponding diagrams in the two BSEs for the charged pions are diquark exchange diagrams. Again, the effects of these
diagrams are suppressed by factors of $m_\pi^2/m_{qq}^2$ and we neglect these due to the heavy mass of the lowest lying
scalar diquark of the order of 800-900 MeV \cite{Eichmann:2016yit,Barabanov:2020jvn}.  

Furthermore, we need to point out a technical difficulty we face in the present calculation. In the meson-exchange 
kernels shown in Fig.~\ref{fig:BSE} (see also Eq.~\ref{eq:BSEkernel_tchannel}) the quark-meson interaction vertex is given 
by the meson BS amplitude. Moreover, in order to access the time-like region, where bound states fulfil $P^2=-M^2$, 
with $P$ the total bound-state momentum and $M$ the bound-state mass, we need to work with complex momenta in the 
BSE, since we work in Euclidean spacetime. This implies, in particular, that the BS amplitudes used as vertices 
need to be evaluated for complex relative quark momenta (i.e. the momenta $p$ in Eq.~\eqref{eq:homogeneousBSE}). Solving BSEs for complex
relative momenta is currently only possible in the RL approximation. Therefore, the BS amplitudes used as vertices 
are obtained from a RL-truncated BSE calculation. Unfortunately, we have no way of estimating at the moment how 
big the quantitative effect of such an approximation on our results is. We do not expect, however, that 
any qualitative statements change. 

Finally, let us point out that neither the gluon exchange nor the meson exchange interaction mechanisms induce any 
isospin breaking on their own. However, if explicit isospin breaking is introduced by a choice of different up- 
and down-quark masses, the meson exchange terms will cease to be degenerate and introduce non-trivial mixing of 
the different BSE amplitudes. Thus isospin breaking occurs on different levels: it results in different dressing functions
for the up- and down-quark propagators but it also results in a flavour dependent QCD interaction manifest in the different
meson exchange terms in our truncation. 

\subsection{\label{sec:qed} Electromagnetically-induced Isospin Breaking}
To describe electromagnetic effects we consider a dressed photon exchange term in the BSE kernel and the corresponding 
term in the quark DSE (see Figs.~\ref{fig:DSE} and \ref{fig:BSE}). The quark-photon interaction is described by the 
fully-dressed quark-photon vertex. Even though the non-perturbative structure of the quark-photon vertex is by now 
reasonably well known \cite{Ball:1980ay,Frank:1994mf,Maris:1999bh,Chang:2010hb,Eichmann:2014qva,Tang:2019zbk}, for 
the qualitative study of this work we will choose a simpler model.

The non-transverse part of the quark-photon vertex is given by the Ball-Chiu construction \cite{Ball:1980ay} in terms 
of the quark dressing functions $A_f(p^2)$ and $B_f(p^2)$, with $S_f^{-1}(p)=-i\Slash{p}A_f(p^2)+B_f(p^2)$ the inverse quark 
propagator. In this work we simplify the calculation further, employing the model used in \cite{Bonnet:2018sbs}, 
where only the $\gamma^\mu$ component of the Ball-Chiu vertex is used. In this way, the QED contribution takes the 
same form as the gluon rainbow-ladder term and can be incorporated quite easily by adding 
\begin{equation}\label{eq:QEDmodel}
\alpha_{\text{QED}}(k^2)= \alpha_0 Q_gQ_f Z_{\text{QED}}(k^2)~,
\end{equation}
to $\alpha_{\text{QCD}}$ in Eq.~\ref{eq:MTmodel}. Here, $\alpha_0 \approx 1/137$ corresponds to the electromagnetic coupling, 
$Q_{f,g}$ are the charges of the two quarks attached to the photon line with flavour $f,g$ and $Z_{\text{QED}}$ the dressing 
of the leading tensorial structure of the quark-photon vertex. It is parametrised by 
\begin{equation}
Z_{\text{QED}}(k^2) = \frac{A_f(k^2) + A_g(k^2)}{2}~,\label{eq:QEDmodel2}
\end{equation}
where $i,j$ represent the two quarks at both ends of the photon. This flavour averaging has no effect on
the self-energy in the quark-DSE, but is appropriate for the kernels in flavour non-diagonal meson states.

Solving the coupled system of DSE and BSE using different masses and charges 
for up and down quarks we are now in a position to study the effects of the different sources of isospin breaking.

\section{\label{sec:results} Results}

\begin{table*}[t]
	\centering
	\begin{tabular}{|l|l|l|l|l|}
		\hline       &             &                 &                  &     \\[-1em]    
		             & $\bar{X}$   &                 &                  & $X$  \\
		  $M$[MeV]   & $m_u = m_d$ & $m_u \neq m_d$  & $m_u = m_d$      & $m_u \neq m_d$  \\
		             & $\alpha_0=0$& $\alpha_0=0$    & $\alpha_0=1/137$ & $\alpha_0=1/137$ \\ \hline\hline
		$\pi^0$      &  134.5      &  132.5          &  136.0           & 133.4             \\ \hline
		$\pi^{\pm}$  &  134.5      &  134.2          &  139.6           & 139.7$^\dagger$    \\ \hline
  $\pi^{\pm}-\pi^0$  & 0           &    1.7          &        3.6       & 6.3                 \\ \hline\hline
		  $K^0$      &  494.7      &  497.5          &     495.2        & 497.7$^\dagger$   \\ \hline
          $K^{\pm}$  &  494.7      &  492.1          &     497.2        & 493.7$^\dagger$    \\ \hline
        $K^0-K^\pm$  & 0           &    5.4          &         -2.0     & 4.0                 \\ \hline\hline
		$\rho^0$     &   720.3     &   721.5         &   721.1          &     721.4            \\ \hline
		$\rho^{\pm}$ &    720.3    &   719.9         &    722.0         &     720.9             \\ \hline
 $\rho^{\pm}-\rho^0$ & 0           &   -1.6          &    0.9           &       -0.5              \\ \hline
	\end{tabular}
		\caption{\label{table:masses}
Pion, kaon and rho masses $M$ for the full system depicted in Figs.~\ref{fig:DSE} and \ref{fig:BSE} 
		with quark masses $m_u = 6.3$~MeV, $m_d =8.1$~MeV, $m_s =82.8$~MeV (third and fifth column), average quark masses 
		$m_{ud}=m_u=m_d=7.2$~MeV (second and fourth column) and electromagnetic isospin breaking switched on and off. 
		A dagger indicates that the quark masses where fitted to obtain those meson masses.}
\end{table*}

First we need to fix the QCD-parameters, the quark masses $m_u$, $m_d$ and $m_s$, and the parameters of our model interaction, 
the Maris-Tandy parameters $\Lambda$ and $\eta$. The model parameter $\Lambda$
sets the scale and is adjusted such that the experimental decay constant of the charged pion is reproduced. 
Furthermore it is well known from pure RL calculations \cite{Eichmann:2016yit} that within a certain range, 
experimental quantities are fairly insensitive to the value of the other model parameter, $\eta$. 
In our calculations we choose $\eta = 1.42$ and $\Lambda = 0.74$. 

This leaves the up-, down- and strange-quark masses to be determined. We adjust 
these such that a complete calculation with rainbow-ladder gluon exchange, meson exchange and photon exchange 
(cf. Figs.~\ref{fig:DSE} and \ref{fig:BSE}) reproduces the experimental masses for the charged pion and the 
neutral and charged kaon. The corresponding results are displayed in the rightmost column of Tab.~\ref{table:masses}. 
We find $m_u = 6.3$~MeV, $m_d =8.1$~MeV and $m_s =82.8$~MeV for the quark masses in a MOM-scheme evaluated at 
a renormalisation scale $\mu = 19$ GeV. All other results for meson masses are then model predictions. Note 
that the masses of the vector meson in general come out too small by about 50 MeV; this is a well-known artefact 
of the Maris-Tandy model and is remedied only in a full calculation taking into account corrections from the 
gluon self-interaction \cite{Fischer:2009jm} and from the decay into two pions\cite{Williams:2018adr}.

Here we like to point out the well-known result that the dynamically generated mass 
function $M_f(p^2)$, {\it cf.} Eq.\ \eqref{QuarkProp}, displays a significant enhancement
of the mass splitting for values of $p^2$ being in the sub-GeV region. Taking into account only the 
flavour-blind gluon-mediated interaction one obtains
\begin{equation} \label{eq:deltaM}
M_u(0) - M_d(0) \approx 5 (m_u-m_d)
\end{equation} 
Including the pion exchange for the two light flavours lowers $M_{u,d}(0)$ both by 37 MeV, 
and consequently the relation \eqref{eq:deltaM} stays valid. The photon exchange
increases $M_u(0)$ by slightly less than one MeV, resp., 0.3 \%, which is of the 
expected size due to $Q_u^2 \alpha_0 =$ 0.003.
Due to the lower absolute value of the $d$ quark charge $M_d(0)$ is increased by approximately 
a quarter MeV (and $M_s(0)$ by half a MeV). Clearly, these small additional mass increases leave 
us more or less with the mass difference \eqref{eq:deltaM}.

The enhancement \eqref{eq:deltaM} from $m_u-m_d = 1.8$ MeV to $ M_u(0) - M_d(0) \approx $ 9 MeV 
is important for an understanding of the induced mass difference of the neutral to the charged kaons
in terms of the quark masses. 
The strong-interaction induced mass difference for the  $K$ mesons is expected to be due to binding 
effects smaller than the mass difference of the two different light valence quarks.
As a matter of fact, from our results for the K mesons, see  Tab.~\ref{table:masses},
we infer that it is smaller than
$ M_u(0) - M_d(0)$ but due to \eqref{eq:deltaM} it is significantly larger than the 
current mass difference $m_u-m_d$.

In order to analyse our results for the meson masses
and in particular the separate effects of isospin breaking due to different up-/down-quark 
masses and the electromagnetic interaction in the context of the available literature we adopt the notation 
of the FLAG review \cite{Aoki:2021kgd}. As described above, the quark masses are then determined by the 
system of equations
\begin{align}
M_{\pi^+}(m_u, m_d, m_s, \alpha_0) &= M_{\pi^+}^{exp},\\
M_{K^+}(m_u, m_d, m_s, \alpha_0)   &= M_{K^+}^{exp},\\
M_{K^0}(m_u, m_d, m_s, \alpha_0)   &= M_{K^0}^{exp},
\end{align}
where the right hand sides are the experimental values and $\alpha_0 = 1/137$ is precise enough for our study.
All calculated masses are functions of $m_u,m_d,m_s,\alpha_0$, and since 
isospin breaking effects are small, in a hadronic 
quantity $X$ they may be described to first order by expansions in $\delta_m$ and $\alpha_0$ with
\begin{align}\label{firstorder}\hspace*{-2mm}
X(m_u, m_d, m_s, \alpha_0) &=\bar{X}(m_{ud},m_s) \nonumber\\
&+ \delta m \,A_X(m_{ud},m_s) \nonumber\\
&+ \alpha_0 \,B_X(m_{ud},m_s)\nonumber\\
&+\delta m \,\alpha_0 \,C_X(m_{ud},m_s) \\
&= \bar{X} + X^{SU(2)} + X^{\gamma} + X^{SU(2),\gamma}. \nonumber
\end{align}
Here, $m_{ud}=1/2(m_u+m_d)$ is the average light-quark mass.
Higher order effects in this expansion are traditionally expected to be small \cite{Aoki:2021kgd}. In order to
gauge this expectation, we also included the lowest order mixed quantity proportional to $\delta m \,\alpha_0$ 
to facilitate our analysis. The size of this additional correction can be estimated from comparing results 
for $X$ and $\bar{X}$ with results obtained from individually setting $\delta m$ or $\alpha_0$ to zero.

Our results for the isospin doublet kaons and the isospin triplet pions and $\rho$-mesons as well as their 
mass differences are shown in Tab.~\ref{table:masses}. We find that the kaon mass splitting is dominated by
strong isospin breaking, $(m_{K^0}  - m_{K^\pm})^{SU(2)} = 5.4$ MeV
with a much smaller electromagnetic breaking with opposite sign
 $(m_{K^0}  - m_{K^\pm})^{\gamma} = - 2.0$  MeV and thus following the expectation derived from  
Eq.\ \eqref{eq:DashenCorr}.
In contrast, the pion mass splitting is dominated by the electromagnetic effect  
$(m_{\pi^\pm}  - m_{\pi^0})^{\gamma} = 3.6$  MeV, and an addition due to the strong isospin breaking
of only half the size, $(m_{\pi^\pm}  - m_{\pi^0})^{SU(2)} = 1.7$  MeV. 

In both cases we also observe that the two sources of breaking
do not add linearly, but there are non-negligible effects due to mixed higher order terms in the expansion 
Eqs.~(\ref{firstorder}). We find $(m_{\pi^\pm}  - m_{\pi^0})^{SU(2),\gamma} = 1.0$  MeV for the pions and
$(m_{K^0}  - m_{K^\pm})^{SU(2),\gamma} = 0.6$ MeV for the kaons. 
The mass of the neutral pion is only slightly affected by isospin breaking, most of the
difference stems from the shift in the mass of the charged pions. The biggest single effect is the 
electromagnetically induced mass increase of the charged pions, and all isospin breaking effects are
adding up.
(NB: It should be noted, however, that our calculation overestimates the pion mass difference, we obtain 
6.3 MeV instead of the observed 4.6 MeV \cite{ParticleDataGroup:2020ssz}.)
For the kaons we note the partial cancellation of electromagnetic and strong isospin breaking effects.
For the charged kaons this cancellation is almost complete, whereas for the neutral kaon
there is a larger mass shift due to the strong interaction as the one due to the electromagnetic interaction.

A somewhat different pattern can be observed for the rho meson masses.
First of all, the observed individual shifts are much 
smaller than the ones for the pions or the kaons, and in addition they
largely cancel each other due to opposite signs. But again, we observe that the two sources of isospin
breaking do not add linearly. 

\section{\label{sec:outlook} Conclusions and outlook}

In this work, we determined the effects of isospin breaking from the strong interaction, via different
masses for the up- and down-quark, and the electromagnetic interaction, via different charges of the up-
and down-quark, on the pseudoscalar meson isospin triplet, strange pseudoscalar meson doublet and the
vector meson triplet. We find that the pion splitting is mainly induced by the different quark charges, 
but we also observe some effects due to different masses and even noticeable mixed, {\it i.e.} non-linear,
effects of both.
The corresponding splitting in the rho triplet is very small which is due to relatively small 
individual mass shifts as well as competing mass and charge splitting effects of opposite sign. 
Again, the non-linear effects are relevant to obtain the resulting mass splitting.
A special role is played by the kaon doublet splitting which results from quite a number of mass 
shifts of similar magnitude but opposite sign.

Our results agree qualitatively with those of other continuum approaches, mainly the $\chi$PT results, 
see, {\it e.g.}, \cite{Bijnens:1996kk,Amoros:2001cp,Bijnens:1997ni}, and with those of lattice QCD
see, {\it e.g.}, \cite{Aoki:2021kgd,Horsley:2015vla,MILC:2018ddw,Fodor:2016bgu,Giusti:2017dmp,Yong:2021qum}. 
In particular, we find that the traditional view that the pion mass difference does not arise from the 
quark mass difference but practically only from the different quark charges is an unjustified 
oversimplification. We also note that our results do not support Dashen's theorem \cite{Dashen:1969eg}
for the electromagnetically induced kaon mass splitting, instead we have
\begin{equation}
\left( m^2_{K^\pm}- m^2_{K^0} \right)_{\mathrm em} \approx 2.0
\left( m^2_{\pi^\pm}- m^2_{\pi^0} \right)_{\mathrm em} \,,
\end{equation}
in agreement with the $\chi$PT result, Eq.~(\ref{eq:DashenCorr}).

\section*{Acknowledgements}
We thank Jan Bonnet and Richard Williams for collaboration and discussions in early stages of this work.  
This work has been supported by Silicon Austria Labs (SAL), owned by the Republic of Austria, the 
Styrian Business Promotion Agency (SFG), the federal state of Carinthia, the Upper Austrian 
Research (UAR), and the Austrian Association for the Electric and Electronics Industry (FEEI). A.S. Miramontes acknowledges CONACYT for financial support.

\bigskip	

\bibliography{isospin}

\end{document}